\documentclass[12pt]{article}
\usepackage{amssymb}
\usepackage{amsmath}
\usepackage{epsfig}

\newcommand{\singlespace}{\renewcommand{\baselinestretch}{1}\large\normalsize}
\newcommand{\doublespace}{\renewcommand{\baselinestretch}{1.6}\large\normalsize}

\newcommand{\beq}{\begin{equation}}
\newcommand{\eeq}{\end{equation}}
\newcommand{\bea}{\begin{eqnarray}}
\newcommand{\eea}{\end{eqnarray}}
\newcommand{\ave}[1]{\langle {#1} \rangle}
\newcommand{\eq}[1]{Eq.~(\ref{#1})}
\newcommand{\eqs}[1]{Eqs.~(\ref{#1})}
\newcommand{\fig}[1]{Fig.~\ref{#1}}
\newcommand{\stt}{s_{22}}
\newcommand{\sff}{s_{55}}
\newcommand{\sss}{s_{77}}
\newcommand{\ptf}{p_{25}}
\newcommand{\pts}{p_{27}}
\newcommand{\pft}{p_{52}}
\newcommand{\pfs}{p_{57}}
\newcommand{\pst}{p_{72}}
\newcommand{\psf}{p_{75}}

\def\dsl{\partial\hspace{-2.5mm}/}
\def\psl{p\hspace{-1.7mm}/}

\def\={\;=\;}
\def\+{\;+\;}

\begin{document}

\begin{center}
\doublespace
\begin{large}
{\bf  NJL-model description of 
Goldstone boson condensation in the color-flavor locked phase} 
\end{large}
\vskip 0.4in

Michael Buballa

{\small{\it Institut f\"ur Kernphysik, TU Darmstadt,
            Schlossgartenstr. 9, 64289 Darmstadt, Germany}}
\end{center}

\vspace{8mm}

\begin{abstract}
A schematic NJL-type model is employed to investigate
kaon and pion condensation in deconfined 
quark matter in the color-flavor locked (CFL) phase, explicitly 
referring to quark degrees of freedom. 
To that end we allow for non-vanishing pseudoscalar diquark condensates 
in addition to the scalar ones which constitute the CFL phase.
Color neutrality is ensured by the appropriate choice of color
chemical potentials. 
The dependence of the free energy in the Goldstone condensed phases 
on quark masses and charge chemical potentials is found to be in
good qualitative -- in most cases also quantitative -- agreement with 
the predictions obtained within the effective Lagrangian approach.
\end{abstract}

\singlespace

\section{Introduction}

It is now generally believed that strongly interacting matter at 
low temperatures and very high densities is a color superconductor 
in the color-flavor locked (CFL) phase~\cite{ARW99,Sho99,Sch00b,EHHS00}
(For reviews on color superconductivity see, e.g., 
Refs.~\cite{RaWi00,Alford,Schaefer,Rischkereview}).
For vanishing quark masses this phase can be characterized by the 
equality of three scalar diquark condensates in the color and flavor 
antitriplet channel,
\beq
    \stt \= \sff \= \sss~,
\label{cfl}
\eeq
where
\beq
    s_{AA'} \= \ave{\,q^T \,C \gamma_5 \,\tau_A \,\lambda_{A'} \,q\,}~.
\label{saa}
\eeq
Here
$q$ is a quark field, and $\tau_A$ and $\lambda_{A'}$,
$A, A' \in \{2,5,7\}$, denote the antisymmetric Gell-Mann matrices 
acting in flavor space and color space, respectively.
In general, these condensates are accompanied by induced color-flavor
sextet condensates, which are, however, small and will be neglected
in this article.

The condensates \eq{cfl} break the original 
$SU(3)_{color} \times SU(3)_L \times SU(3)_R$
symmetry of QCD (in the chiral limit) down to a residual
$SU(3)_{color+V}$, corresponding to a common (``locked'') rotation
in color and flavor space. As a consequence of the breaking of the
color symmetry, all eight gluons receive a mass, 
while the breaking of chiral symmetry leads to the emergence of
eight pseudoscalar Goldstone bosons. 
The latter reflect the fact that there is a continuous set of degenerate
ground states which can be generated from the CFL ansatz, \eq{cfl},
via axial flavor transformations
\beq
    q \;\longrightarrow\; \exp(i\theta_a\frac{\tau_a}{2}\gamma_5)\,q~,\qquad
    a = 1, \dots, 8~.
\label{axial}
\eeq
Under these transformations the scalar diquark condensates, \eq{saa},
are partially rotated into pseudoscalar ones,
\beq
    p_{AA'} \= \ave{\,q^T \,C \,\tau_A \,\lambda_{A'} \,q\,}~.
\eeq
These condensates are expected to become relevant when we consider the 
less perfect but more realistic situation where chiral symmetry is explicitly
broken through non-zero quark masses or external charge chemical potentials.
In this case the CFL state can become unstable against developing non-zero
values of $p_{AA'}$, corresponding to pion or kaon 
condensation~\cite{Sch00c,BS02,KR02}.
 
This behavior has been predicted within low-energy effective field theories 
which can be constructed systematically for excitations much smaller than
the superconducting gap 
$\Delta$~\cite{Sch00c,BS02,KR02,SoSt00,CaGa99,CGN01}.
Basically, this is Chiral Perturbation Theory ($\chi$PT) with the essential 
difference to the well-known $\chi$PT at low densities that at high densities 
the interaction is weak and the coefficients can be calculated from 
QCD within High Density Effective Field Theory
\cite{BS02,Hong00,BBS00}, which is valid for
energies much smaller than the chemical potential. 

Let us briefly summarize the leading-order results which we need for later 
comparison. Adopting the notation of Ref.~\cite{KR02},
the effective meson masses are given by
\beq 
    M_\pi^2 \= 2a\,m_q\,m_s, \quad
    M_K^2 \= a\,(m_q+m_s)\,m_q, 
\label{masses}
\eeq
where we have assumed equal masses for the light quarks, $m_u = m_d =: m_q$.
At asymptotic densities, the coefficient is given by
$a = 3\Delta^2/(\pi^2 f_\pi^2)$~\cite{SoSt00}, where $\Delta$ is the CFL gap 
and the pion decay constant is given by
$f_\pi^2 = (21 - 8\ln 2)\mu^2/ (36\pi^2)$.

The mesons experience effective chemical potentials
\beq
     \tilde \mu_{\pi^+} = \mu_Q~,\quad
     \tilde \mu_{K^+} = \mu_Q + \frac{m_s^2-m_q^2}{2\mu}~,\quad
     \tilde \mu_{K^0} = \frac{m_s^2-m_q^2}{2\mu}~,
\label{mutilde}
\eeq
and the same with the opposite signs for $\pi^-$, $K^-$, and $\bar K^0$.
The above expressions imply that meson condensation takes place if
$\tilde\mu_i$ 
exceeds the mass of the corresponding meson. In this case the 
thermodynamic potential of the system is lowered by 
\beq
     \delta\Omega_i \=  -\frac{f_\pi^2}{2}\,\tilde\mu_i^2\,
     (1 - \cos\theta)^2~, \qquad   \cos\theta \= \frac{M_i^2}{\tilde \mu_i^2}~,
\label{delom}
\eeq
while for $\tilde\mu_i^2 < M_i^2$ the CFL state is stable, i.e., 
$\theta = 0$.

Note that the above expressions are of order zero in $\alpha_s$. 
They are therefore universal in the sense that they should not only hold in
QCD, but in any model with the same symmetry pattern, as long as 
higher-order corrections in the interaction are small.  

Aim of the present article is to study this mechanism and the corresponding
CFL $+$ Goldstone phases within a schematic NJL-type model, explicitly 
referring to the diquark condensates $s_{AA'}$ and $p_{AA'}$.
Our main motivation is the fact that 
NJL-type models have successfully
been used to investigate various color superconducting phases and their
competition (see Ref.~\cite{buballa} and references therein),
but so far the possibility of Goldstone boson condensation has not been
taken into account\footnote{For a recent NJL-model analysis of pion and kaon 
condensation in color non-superconducting quark matter, see
Ref.~\cite{BCPR04}.}. 
In particular the recently proposed gapless phases at densities below the
CFL regime~\cite{AKR04,RSR04,FKR04} 
have only been studied within NJL-type models which neglect 
Goldstone condensates.   
In the present paper we want to lay a basis for a more complete 
analysis in the future. 
We thereby focus on the general mechanism. We investigate whether and
under which conditions Goldstone condensation takes place in
NJL-type models and compare the results 
with the predictions of the effective theories. 
To keep the analysis as simple as possible we do not consider 
$\bar q q$ and $\bar q i\gamma_5 \tau_a q$ condensates at the present 
stage. An extension of the model in this direction is straight forward.

\section{Chiral transformations}
\label{chiral}

Before we define our model we should discuss the 
axial transformations introduced in \eq{axial} in some more details.
In this article we consider exclusively flavor non-diagonal 
modes of \eq{axial}.
According to their quantum numbers, 
these modes can be identified with charged pions (``$\pi^\pm$'', $a = 1,2$), 
charged kaons (``$K^\pm$'', $a = 4,5$), 
or neutral kaons (``$K^0$'', $a = 6,7$). 

As mentioned earlier, under these transformations the scalar diquark 
condensates are partially rotated into pseudoscalar ones. 
For a $K^0$ mode, for instance, the transformed condensates read 
\begin{alignat}{2}
K^0: \qquad\qquad \stt' &\= \cos\frac{\theta}{2}\;\stt  \qquad & \qquad
                  \pft' &\=  i\,\sin\frac{\theta}{2}\;
                  (\hat\theta_6 \,-\, i\hat\theta_7)\;\stt
\nonumber \\
               \sff' &\= \cos\frac{\theta}{2}\;\sff  \qquad & \qquad
               \ptf' &\=  i\,\sin\frac{\theta}{2}\;
               (\hat\theta_6 \,+\, i\hat\theta_7)\;\sff
\nonumber \\
               \sss' &\= \sss  &&
\label{K0}
\end{alignat}
where $\theta = \sqrt{\theta_6^2 + \theta_7^2}$ and
$\hat\theta_a = \theta_a/\theta$. 
The primed and unprimed condensates refer to the values in the 
transformed state and in the original CFL state, respectively.  
Similarly, under $K^\pm$ transformations, $\stt$ and $\sss$ are partially 
rotated into $\pst$ and $\pts$, respectively, while $\pi^\pm$ transformations,
rotate $\sff$ into $\psf$ and $\sss$ into $\pfs$.

For equal quark masses, because of the residual $SU(3)_{color+V}$ symmetry 
of the CFL state, 
the same combination of diquark condensates could be reached via 
axial {\it color} transformations
\beq
    q \;\longrightarrow\; 
    \exp(i\theta_a\frac{\lambda_a^T}{2}\gamma_5)\,q~,\qquad
    a = 1, \dots, 8~.
\label{axialc}
\eeq
For instance, for $a = 6,7$, this transformation yields 
\begin{alignat}{2}
K^{'0}: \qquad\qquad \stt' &\= \cos\frac{\theta}{2}\;\stt  \qquad & \qquad
                  \ptf' &\=  i\,\sin\frac{\theta}{2}\;
                  (\hat\theta_6 \,+\, i\hat\theta_7)\;\stt
\nonumber \\
               \sff' &\= \cos\frac{\theta}{2}\;\sff  \qquad & \qquad
               \pft' &\=  i\,\sin\frac{\theta}{2}\;
               (\hat\theta_6 \,-\, i\hat\theta_7)\;\sff
\nonumber \\
               \sss' &\= \sss  &&
\label{K0p}
\end{alignat}
Hence, as long as \eq{cfl} holds exactly, both transformations,
\eq{K0} and \eq{K0p}, lead to the same state.
On the other hand, in the more realistic case of unequal quark masses, 
there is no exact flavor $SU(3)_V$ to begin with and, thus, 
no exact $SU(3)_{color+V}$ in the CFL phase. 
In this case the scalar diquark condensates $\stt$, $\sff$, and $\sss$ are 
in general not equal and the results of an axial flavor transformation and
the corresponding axial color transformation will be different.
Therefore, in order to distinguish between these two transformations,
we indicate axial color transformations by a prime, i.e., $\pi^{'\pm}$, 
$K^{'\pm}$, and $K^{'0}$.

Looking at \eqs{K0} and (\ref{K0p}) one might think that a maximally
meson condensed state corresponds to a transformation angle
$\theta = \pi$, i.e., when two of the three scalar diquark condensates
are rotated away completely, and the two pseudoscalar diquark condensates
receive their maximum value. This is, however, not true. 
To see this, it is instructive to perform an axial transformation on
an idealized vacuum state with 
$\ave{\bar uu} = \ave{\bar dd} = \ave{\bar ss} =: \phi_0$.
For a $K^0$ transformation, as defined above, this state goes over 
into
\begin{alignat}{1}
    &\ave{\bar uu}' \= \phi_0~,  \qquad
    \ave{\bar dd}' \= \ave{\bar ss}' \= \cos\theta\;\phi_0~,  \qquad
    \ave{\bar q\, i\gamma_5 \tau_a q}' \= \sin\theta\;\hat \theta_a\;\phi_0,
    \quad a = 6,7~.
\label{K0qbarq}
\end{alignat}
Obviously, the condensate with the quantum numbers of a kaon is maximal
at $\theta = \pi/2$. 
Therefore, since in a color superconductor quark-antiquark states and
diquark states can mix, maximally meson condensed states should  
correspond to $\theta = \pi/2$ in this case as well. 
This is consistent with the effective Lagrangian 
description\footnote{The angle $\theta$ in \eq{delom} corresponds to 
the rotation angle $\pi^a/f_\pi$ of the
chiral field $\Sigma = \exp(i \pi^a \lambda^a/f_\pi)$ which transforms
like $\Sigma  \rightarrow L\Sigma R^\dagger$, where $L$ and $R$ are the
$SU(3)$ matrices which transform the left and right handed quarks:
$q_L \rightarrow L\,q_L$, $q_R \rightarrow R\,q_R$. Hence, for 
$L = R^\dagger$, the transformation angle for the quarks is $\theta/2$,
which is consistent with \eq{axial}.}.
We will see that it is also confirmed by our numerical results.

\section{Model}

We consider an NJL-type Lagrangian with a point-like color-current
interaction,
\beq
    {\cal L} \= \bar q\,(i\dsl - \hat m)\,q 
    \;-\; g\,\sum_{a=1}^8\;(\bar q \gamma^\mu \lambda_a q)^2~,
\label{lagrangian}
\eeq
where $q$ denotes a quark field with three flavor and three color degrees
of freedom, $\hat m = diag_f(m_u, m_d, m_s)$, and $g$ is a dimensionful
coupling constant. 

Performing a Fierz transformation into the particle-particle channel,
the interaction part can be rewritten as
\beq
{\cal L}_\mathit{qq} \;=\;\frac{2}{3}\,g\,\sum_{A,A'}
\Big[ (\bar q i\gamma_5C \tau_A\lambda_{A'}\bar q^T)
       (q^T C i\gamma_5 \tau_A\lambda_{A'} q) +\;  
       (\bar q C \tau_A\lambda_{A'}\bar q^T)
       (q^T C  \tau_A\lambda_{A'} q) \Big] \+ \dots~,
\eeq
where we have listed only those terms which could potentially lead to 
a condensation in the channels $s_{AA'}$ or $p_{AA'}$. 
The mean-field thermodynamic potential in the presence of these condensates
is then given by 
\beq
    \Omega(T,\{\mu_{fc}\}) \;=\; -T \sum_n \int \frac{d^3p}{(2\pi)^3} \;
    \frac{1}{2}\,{\rm Tr}\; \ln \Big(\frac{1}{T}\,S^{-1}(i\omega_n, \vec p)
    \Big)
    \+ \frac{2}{3}\,g\, \sum_{A,A'} (|s_{AA'}|^2 \,+\,|p_{AA'}|^2)
\label{Omega}
\eeq
where $T$ is the temperature, $\omega_n = (2n-1)\pi T$ are fermionic 
Matsubara frequencies,
and $\{\mu_{fc}\}$ ($f \in \{u,d,s\}$,
$c \in \{r,g,b\}$) denotes the set of chemical potentials related to the 
conserved flavor and color densities $n_{fc}$ of the Lagrangian.
They are often given as linear combinations of the chemical potentials
$\mu$, $\mu_Q$, $\mu_3$, and $\mu_8$~\cite{SRP02}, which are related to the 
total
quark number density $n$, the electric charge density $n_Q$, and the
color densities $n_3 = n_r - n_g$ and $n_8 = (n_r + n_g - 2n_B)/\sqrt{3}$,
respectively ($n_c := \sum_f n_{fc}$). 

The inverse propagator in Nambu-Gorkov formalism reads
\beq
    S^{-1}(p) \= \left(\begin{array}{cc} \psl + \hat\mu\gamma^0 - \hat m 
    & \sum_{AA'}(\Delta^s_{AA'}\gamma_5 + \Delta^p_{AA'})\,\tau_A \lambda_{A'}
\\
      \sum_{AA'}( -\Delta^{s\;*}_{AA'} \gamma_5 + \Delta^{p\;*}_{AA'})\, 
      \tau_A \lambda_{A'}
    & \psl - \hat\mu\gamma^0 - \hat m \end{array}\right)~,
\label{NGSinv}
\eeq
with $\hat\mu = diag_{fc}(\mu_{fc})$ and the scalar and pseudoscalar
diquark gaps
\beq
    \Delta^s_{AA'} \= -\frac{4}{3}\;g\;s_{AA'}~, \quad 
    \Delta^p_{AA'} \= -\frac{4}{3}\;g\;p_{AA'}~.
\eeq
In the following we restrict ourselves to the three scalar condensates
of the CFL ansatz, \eq{cfl}, plus two pseudoscalar condensates,
corresponding to either $\pi^\pm$ ($p_{57}$ and $p_{75}$), 
$K^\pm$ ($p_{27}$ and $p_{72}$), or $K^0$ modes ($p_{25}$ and $p_{52}$).
In other words, we allow for condensation only in one of these modes
at a time, which is a reasonable assumption~\cite{KR02}.
The inverse propagator, which is a $72\times 72$ matrix, can then be
decomposed into a $40\times 40$ block and two $16\times 16$ blocks. 
Making use of the fact that the integrand in \eq{Omega} does not depend
on the direction of the 3-momentum, the dimensionality of the blocks 
can further be reduced by a factor one half. 
The determinants of the remaining $20\times 20$ and $8\times 8$ matrices 
and the Matsubara sum are evaluated numerically. 

Until this point, the thermodynamic potential depends on our choice of the 
diquark gaps. 
As standard, the stable selfconsistent solutions
are given by the values of $\Delta^s_{AA'}$ and $\Delta^p_{AA'}$ which
minimize $\Omega$. 
In most cases below, however, we will restrict the minimization 
procedure to the subspace of condensates which can be obtained from the
CFL solution ($\Delta^p_{AA'}=0$) via chiral rotations, \eq{axial} or
\eq{axialc}, in order to compare our results with the effective Lagrangian
approach.

\section{Numerical results}

In most of our numerical studies we adopt the model parameters of 
Ref.~\cite{BuOe02}, namely a 3-momentum cut-off $\Lambda = 600$~MeV
and $g\Lambda^2 = 2.6$ for the coupling constant.
For comparison with the predictions of the effective Lagrangian approach
we keep the quark masses as variable parameters.
As before, we assume equal masses for up and down quarks,
$m_u = m_d =: m_q$.
All calculations are performed at zero temperature and at 
fixed $\mu = 400$~MeV. 

To have a well-defined starting point, we begin our analysis in the 
chiral limit, $m_q = m_s = 0$ and with equal chemical potentials for 
all quarks, i.e., $\mu_Q = \mu_3 = \mu_8 = 0$. 
For this case we find a CFL solution with 
$\Delta_{22}^s = \Delta_{55}^s = \Delta_{77}^s = 104.2$~MeV. 
As expected, this is not a unique solution for the ground state,
but an infinite set of degenerate ground states can be obtained via chiral 
rotations.
In particular, any rotations in $\pi^\pm$, $K^\pm$, or $K^0$ direction
leave the ground state free energy invariant. 

Next we introduce a finite strange quark mass $m_s = 120$~MeV,
while leaving the up and down quarks massless. 
The explicit breaking of the $SU(3)$ symmetry causes a slight 
asymmetry in the ground state, i.e., instead of an ideal CFL solution,
\eq{cfl}, we get a 5\% splitting of the gap parameters: 
$\Delta_{22}^s = 105.4$~MeV 
and $\Delta_{55}^s = \Delta_{77}^s = 100.7$~MeV.
It turns out that the corresponding free energy is still
invariant under chiral rotations into the $\pi^\pm$ direction, 
whereas it becomes {\it disfavored}, if we perform a rotation into 
the kaonic directions.
At first sight, this result seems to contradict the effective Lagrangian 
results, as summarized in \eqs{masses} - (\ref{delom}).

At this point we should notice that, by construction, the effective Lagrangian 
contains only colorless states (as a result of integrating out static 
gluons, see, e.g.,Ref.~\cite{CaGa99})
whereas the above ground state of the NJL model is not color neutral. 
In fact, it is well known that the introduction of unequal
masses in the CFL phase leads to colored solutions in NJL-type models,
unless this is corrected for by introducing appropriate color chemical 
potentials~\cite{SRP02}. 
This can be thought of as an effective way of simulating static
gluon background fields. 
In the above example we need to introduce a chemical potential 
$\mu_8 = -10.12$~MeV in order to obtain a color neutral CFL solution.
This solution has a smaller splitting of the gaps, 
$\Delta_{22}^s = 103.4$~MeV  and $\Delta_{55}^s = \Delta_{77}^s = 101.8$~MeV,
and is about 0.2~MeV/fm$^3$ higher in free energy as the colored one.

\begin{figure}[h!]
\begin{center}
\epsfig{file=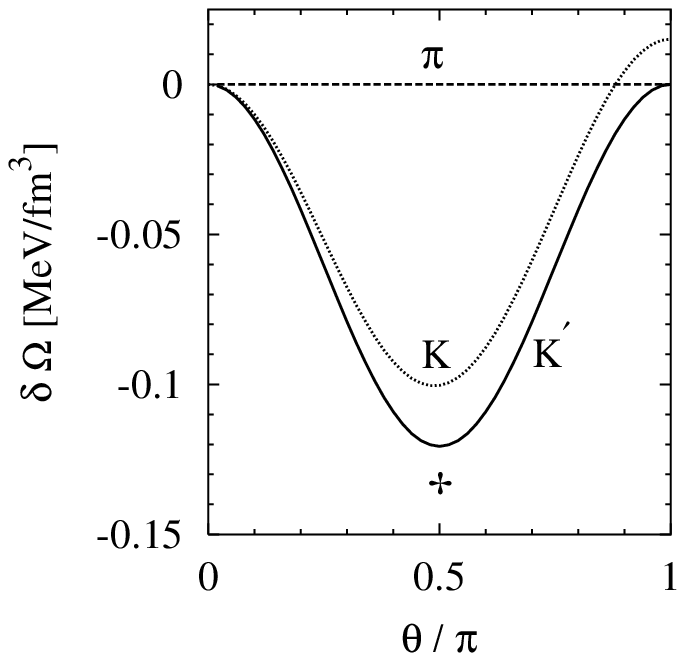,width = 6.5cm}\hspace{10mm}
\epsfig{file=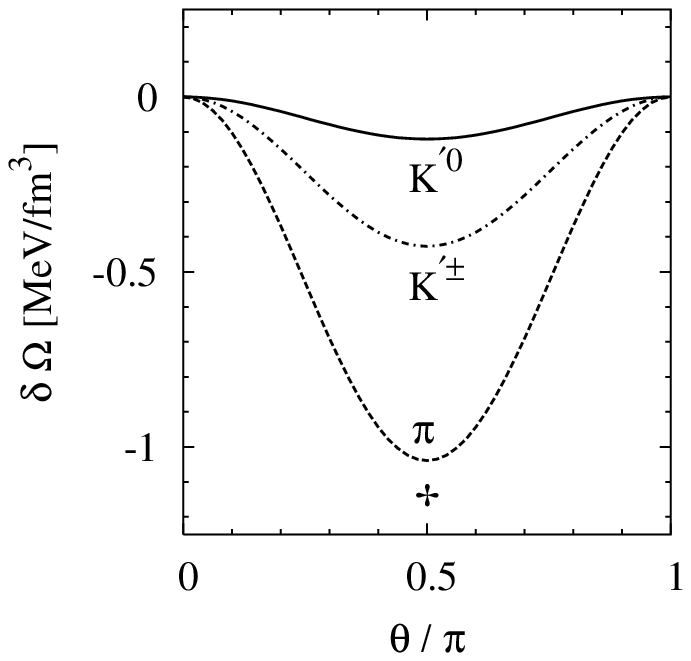,width = 6.5cm}
\caption{\small Relative change of the free energy density as a function 
of the chiral angle for $m_u = m_d = 0$ and $m_s = 120$~MeV.
The calculations have been performed at $\mu = 400$~MeV and 
$\mu_Q = 0$ (left) or $\mu_Q = -50$~MeV (right). 
The color chemical potentials have been adjusted to obtain 
color neutral solutions.
The various lines correspond to different modes as indicated in the
figure. 
The crosses indicate the free energy obtained by minimizing the 
thermodynamic potential for color neutral matter without
restriction to chiral transformations.}
\label{fig1}
\end{center}
\end{figure}

However, the essential observation is that, restricting ourselves to 
color neutral solutions, the CFL state
does no longer correspond to the ground state of the system if we allow
for chiral rotations. This is illustrated in the left panel of \fig{fig1} 
where the relative change of the free energy density is plotted as
a function of the angle $\theta$ for chiral rotations.
While the rotations in pion direction (dashed line) still leave 
the free energy invariant, the kaonic transformations now lead to a reduction 
of $\Omega$. (Note that charged and neutral kaon solutions are degenerate
under the present conditions.)
Here the dotted line indicates the result of an axial flavor 
rotation ($K^\pm$ or $K^0$), and the solid line the result of an axial color 
rotation ($K^{'\pm}$ or $K^{'0}$). 
As one can see, the latter is more favored and symmetric 
about a minimum at $\theta = \pi/2$, while the former is less favored and
slightly asymmetric. 

This can be explained by the fact that, e.g., a $K^0$ 
rotation transforms the larger $\stt$, in which $u$ and $d$
quarks are paired, into the $us$ condensate $\pft$, whereas
the smaller $us$ condensate $\sff$ is rotated into the $ud$
condensate $\ptf$. Hence, the number of strange quarks is larger 
at $\theta = \pi$  than at  $\theta = 0$, which explains why  
$\theta = \pi$ is less favored. 
In a $K^{'0}$ transformation, on the other hand, $ud$ condensates
remain $ud$ condensates and $ds$ condensates remain $ds$ condensates,
because axial color transformations do not change the flavor structure. 
Therefore the number of
strange quarks at $\theta = \pi$ is equal to the number of strange 
quarks at $\theta = 0$, and both states are degenerate. 

None of the above results corresponds
to the real minimum of the thermodynamic potential, because we have 
restricted ourselves to certain chiral rotations of the CFL solution 
without allowing for ``radial'' variations. 
This means, instead of freely varying the scalar and pseudoscalar 
condensates under consideration, we have linked them to the CFL solution
and a single parameter $\theta$, e.g., as given in \eqs{K0} and 
(\ref{K0p}) for the $K^0$ or the $K^{'0}$ case, respectively.
If we abandon this constraint we find solutions with even lower free 
energies, indicated by the cross in the figure. 
These solutions are very similar to a $K'$ state at $\theta = \pi/2$, in the 
sense that, e.g., $|\Delta^s_{22}| = |\Delta^p_{25}|$ and  
$|\Delta^s_{55}| = |\Delta^p_{52}|$, but the values of the $ud$ gaps
are slightly larger (74.2~MeV = 104.9~MeV/$\sqrt{2}$) and the values
of the $us$ gaps are slightly smaller (70.9~MeV = 100.2~MeV/$\sqrt{2}$)
than the rotated CFL gaps.

In fact, in our context the $K'$-transformations should also be 
interpreted as certain ``radial'' transformations on top of the
axial flavor transformation $K$, and not as axial color transformations.
Note that, even in the chiral limit, 
axial color transformations, \eq{axialc}, are neither a symmetry of QCD 
nor of our NJL-model Lagrangian, \eq{lagrangian}.  
They are just a symmetry of the mean-field thermodynamic potential, as 
long as we restrict ourselves to the given set of diquark condensates. 
It will no longer be the case if, e.g., $\bar qq$ condensates are
taken into account.  
In the present model, the difference between the $K$ and $K'$ 
transformations is due to the non-equality of 
the diquark condensates in the CFL phase (cf. \eqs{K0} and (\ref{K0p})). 
This is a higher-order effect, 
which is not taken into
account in the leading-order results of the effective Lagrangian
approach. 
Hence $K$ and $K'$ transformations are in principle equally good starting 
points for a comparison with the effective theory.
However, in some aspects the $K'$ transformations behave more similar
to the kaon modes in the effective theory since they are symmetric 
about $\theta = \pi/2$\footnote{For instance, the color chemical potentials 
needed to neutralize the $K'$ solution at the minimum at $\theta=\pi/2$ 
satisfy the relation $\mu_3 = \sqrt{3}\,\mu_8$. This is in nice agreement 
with the ratio between the third and the eighth color component of the 
static gluon field, $\phi_3^c = \sqrt{3}\,\phi^c_8$,
obtained in Ref.~\cite{Kry03} for neutral CFL $+$ $K^0$ solutions.
This relation does not hold for the $K$ solution in our model.}.
Moreover, they come closer to the true minimum of the thermodynamic
potential. 
Therefore, for most of quantitative studies in this article we 
perform $K'$ transformations rather than $K$ transformations. 
Note that for the pion modes there is no difference anyway.

In the right panel of \fig{fig1} we show the result of a similar
analysis, again for $m_s = 120$~MeV, but now for a non-vanishing
electric charge chemical potential $\mu_Q = -50$~MeV. 
The color chemical potentials $\mu_3$ and $\mu_8$ have again been
adjusted to ensure color neutrality at each point.  
Under these conditions all flavored Goldstone modes lead to a reduction
of the free energy, but the pionic mode (dashed line) is the most favored 
one, followed by the charged kaons (dash-dotted). The neutral kaon is not
sensitive to $\mu_Q$ and shows the same behavior as in the left panel.  
As before, none of the modes corresponds to the real minimum of the 
thermodynamic potential one gets if the condensates are not constrained to 
certain chiral rotations. This free energy is again indicated 
by a cross in the figure.

\begin{figure}[t]
\begin{center}
\epsfig{file=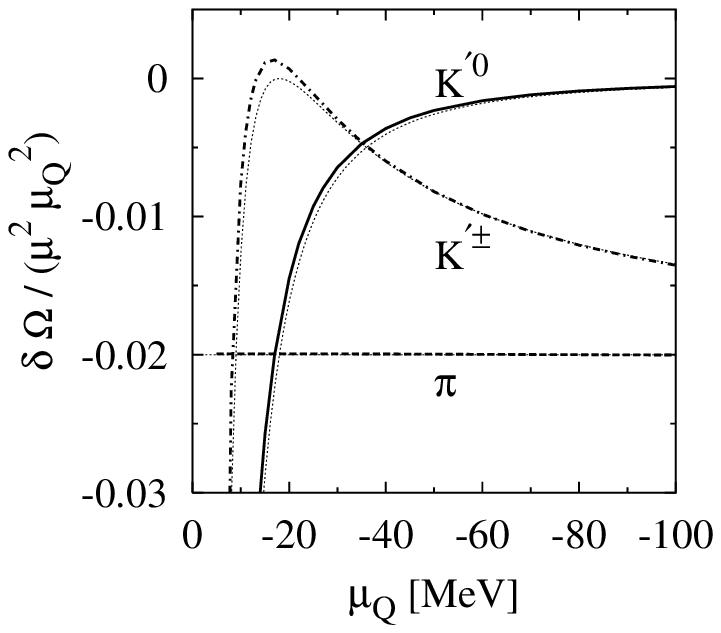,width = 7.0cm}
\quad
\epsfig{file=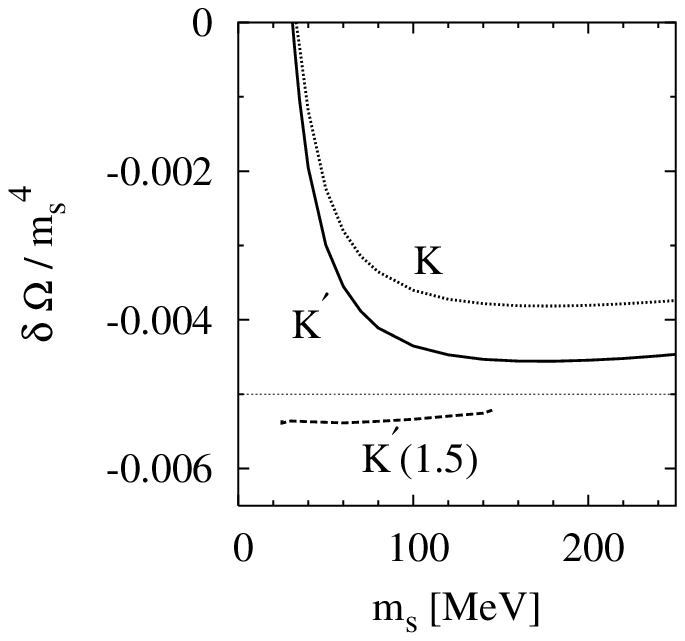,width = 7.0cm}
\caption{\small Free energy density difference between meson condensed
phase ($\theta = \pi/2$) and CFL phase ($\theta = 0$)
as functions $\mu_Q$ (left) or $m_s$ (right). 
The calculations have been performed for $m_u = m_d = 0$ and 
at fixed $\mu = 400$~MeV.
The color chemical potentials have been adjusted to obtain color neutral 
solutions.
In the left panel $m_s = 120$~MeV, in the right panel $\mu_Q = 0$. 
The various lines correspond to different modes as indicated in the
figure. 
The dashed curve labeled ``K'(1.5)'' in the right figure has been
calculated with a reduced coupling constant $g\Lambda^2 = 1.5$.
The bold lines have been obtained within the present model while the
thin dotted lines correspond to \eqs{masses} to (\ref{delom})
with $f_\pi = 80$~MeV.}
\label{fig2}
\end{center}
\end{figure}

At least qualitatively, the results presented in \fig{fig1} are in good
agreement with the predictions of the effective Lagrangian approach. 
For a more quantitative 
comparison we again restrict ourselves to chiral transformations of
the CFL solution. 
In \fig{fig2} we show the behavior of the free energy gain
of the meson condensed solutions ($\theta = \pi/2$) for massless up and down 
quarks as functions of $\mu_Q$ (left panel) and $m_s$ (right panel). 
At each point, the color chemical potentials $\mu_3$ and $\mu_8$ have
been tuned to ensure color neutrality.
 
The left panel shows the behavior of $\delta\Omega$ for the three
Goldstone boson condensates $\pi^\pm$, $K^{'\pm}$, and $K^{'0}$
at fixed chiral angle $\theta = \pi/2$
as functions of $\mu_Q$ at fixed $m_s = 120$~MeV.
Since, according to \eqs{masses} to (\ref{delom}), we expect 
$\delta\Omega_\pi$ to behave like $\mu_Q^2$,  
we plot the dimensionless ratio $\delta\Omega/(\mu^2\mu_Q^2)$.
Indeed, our result for the pionic mode is constant to a very high degree 
(dashed line). 
$\delta\Omega_{K^{'0}}$, on the other hand, does not -- and should not -- 
depend on $\mu_Q^2$, and hence the corresponding curve (solid line) decreases 
like $1/\mu_Q^2$ in the figure. 
Finally, the charged kaons behave more complicated. According to 
\eq{mutilde}, $\mu_{K^\pm}$ vanishes at $\mu_Q = - m_s/(2\mu)$, leading 
to a maximum in the free energy. Qualitatively, this is well 
reproduced in our model calculation (dash-dotted) line. In fact, in the 
$\mu_Q$ interval between -13~MeV and -22~MeV, $\delta\Omega_{K^{'\pm}}$ 
even gets slightly positive. This seems to indicate that the
kaon is not exactly massless but has a small mass of about 5~MeV.
Then, since the leading-order ($\alpha_s^0$) effective theory 
predicts $m_K = 0$ for $m_q = 0$ (see \eq{masses}), this must be a 
higher-order effect due to interactions. There are, for instance, 
QCD corrections proportional to $\alpha_s m_s^2$ 
\cite{KKS04}, which have, however, the wrong sign.
Nevertheless, there could be other terms which lead to the
observed behavior. We will come back to this issue below.

For comparison, we also show the effective Lagrangian results 
for the free energy as given by \eqs{masses} to (\ref{delom}) (dotted lines). 
Here we treat $f_\pi$ as a free 
parameter which is fitted to reproduce the pion. We obtain $f_\pi = 80$~MeV, 
which also gives reasonable fits to the kaons.
The fitted value agrees quite well with the leading-order result
$f_\pi= \sqrt{(21 - 8\ln 2)}\,\mu/ (6\pi) = 83.4$~MeV. 
Again, there should be corrections to this formula due to the interaction.
These could depend on the meson channel as well as on $m_s$ and
$\mu_Q$. 
Of course, meson masses and decay constants could be calculated explicitly
within the NJL model, summing up 
$q\bar q$ loops in Nambu-Gorkov formalism (which are essentially
diquark loops) and coupling them to an external axial current.
This is left for future work.

In the right panel of \fig{fig2} the free energy of the kaonic modes 
($K$ and $K'$) at $\mu_Q = 0$ is shown as a function of $m_s$.
For comparison we show again the result corresponding to  \eqs{masses} to 
(\ref{delom}) for $f_\pi = 80$~MeV (dotted line). 
Since according to these equations we expect 
$\delta\Omega_K$ to behave like $m_s^4$ we plot the ratio 
$\delta\Omega/m_s^4$.
However, whereas for $m_s \gtrsim 100$~MeV this ratio indeed is 
roughly constant, we find strong deviations at lower strange quark 
masses.  In particular, below $m_s = 31$~MeV the kaon modes are no longer 
favored. 

Again, this indicates the presence of higher-order interaction effects.
Note that the deviations from the $m_s^4$ behavior show up in a 
regime where $m_s$  is smaller than the CFL gap $\Delta$.
Hence, terms of the order $m_s^2 \Delta^2$, $\Delta^4$
or $m_s^2 \Delta^4/\mu^2$~\cite{KKS04},
which are exponentially suppressed in the weak-coupling regime,
could be parametrically comparable to $m_s^4$. 
To check this point, we redo the calculation with a
reduced coupling constant, $g\Lambda^2 = 1.5$.
The CFL gaps are then about 30~MeV, 
instead of $\sim 100$~MeV which we had before. 
The resulting $\delta\Omega$ for the $K'$ mode is indicated by the
dashed line in the figure. 
Obviously, this curve shows an almost perfect $m_s^4$ behavior down to
$m_s = 25$~MeV and there is no indication that kaon condensates become
disfavored at small $m_s$. (Unfortunately, we cannot go to even smaller
values of $m_s$ because the numerical uncertainties become too large.) 
This means, there is at least numerical evidence
that the effect seen with our original parameters is indeed
due to interactions.

\begin{figure}[t!]
\begin{center}
\epsfig{file=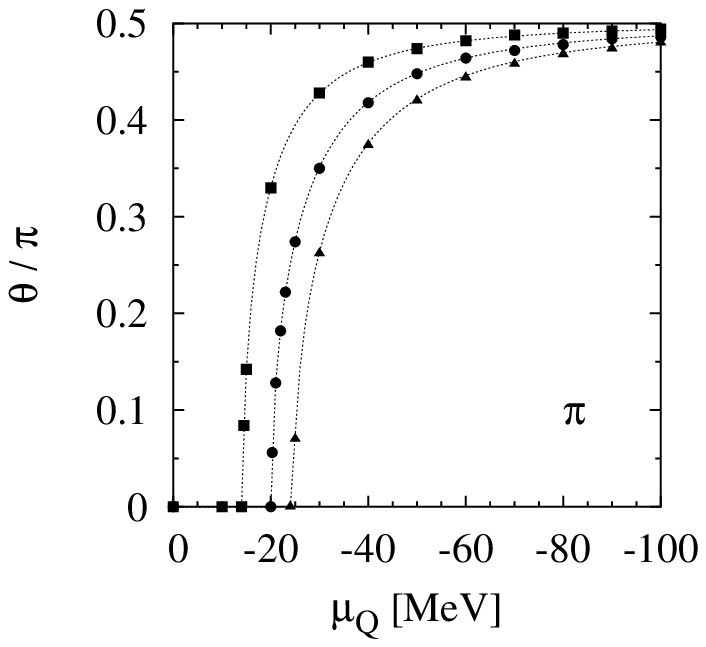,width = 7.0cm}\quad
\epsfig{file=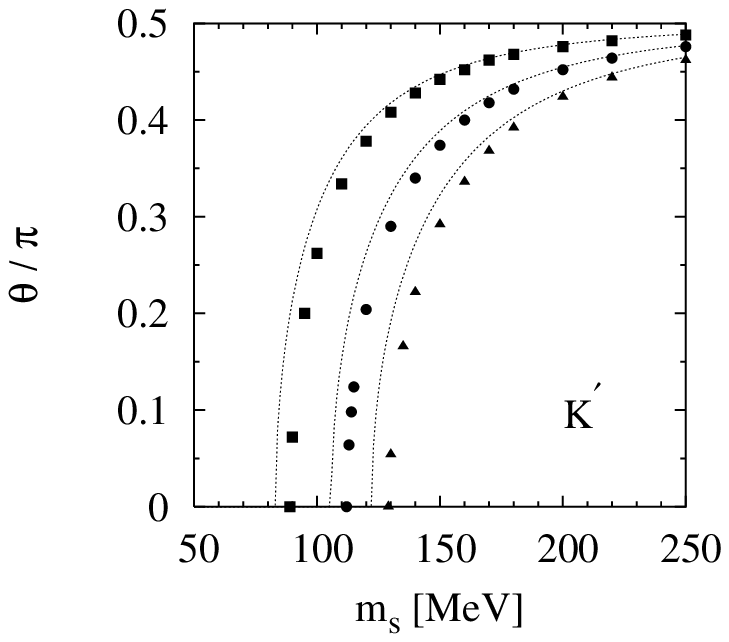,width = 7.23cm}
\epsfig{file=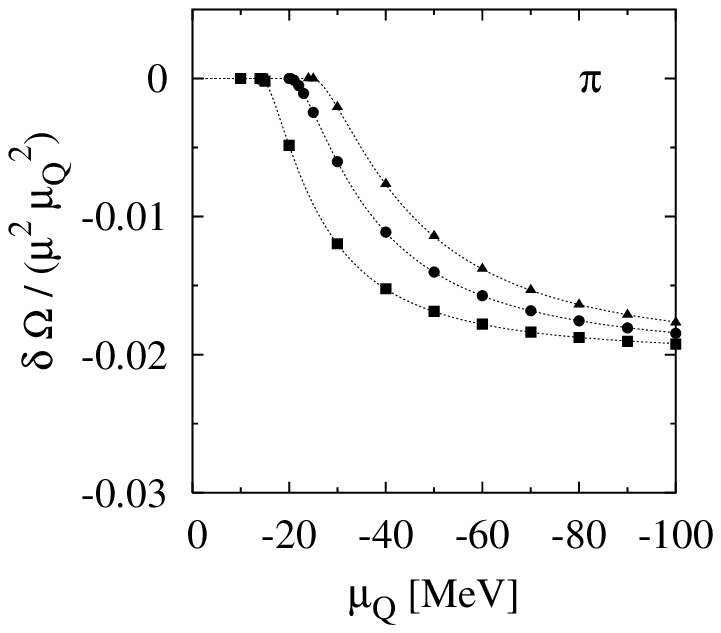,width = 7.0cm}\quad
\epsfig{file=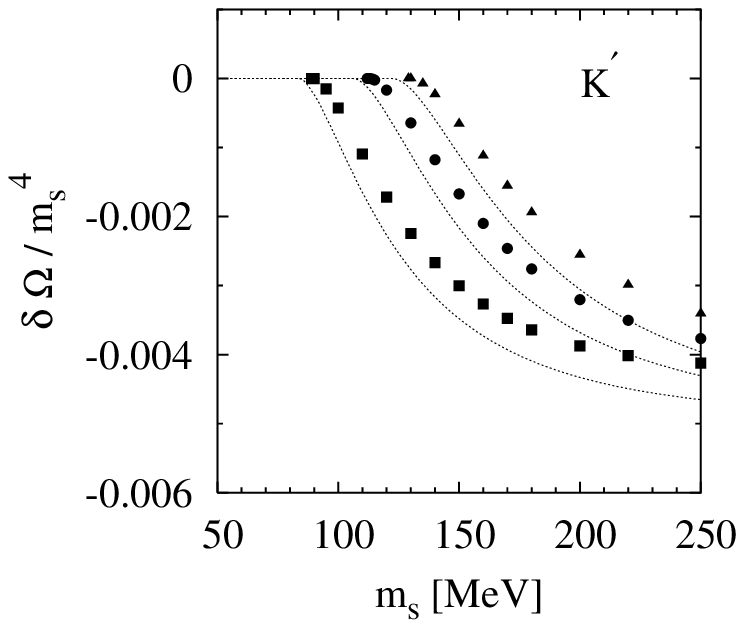,width = 7.23cm}
\caption{\small Chiral angle (upper panels) and corresponding free 
energy gain (lower panels) of meson condensed phases at $\mu = 400$~MeV for 
non-vanishing light quark masses $m_q$.
The full squares, circles, and triangles indicate the NJL model results
for $m_q =$~5~MeV, 10~MeV, and 15~MeV, respectively. 
The dotted lines are the corresponding results of \eqs{masses} to 
(\ref{delom}) with $a = 0.169$ and $f_\pi = 80$~MeV.
Left: pion condensate for $m_s = 120$~MeV and varying $\mu_Q$.
Right: kaon condensate ($K'$) for $\mu_Q = 0$ and varying $m_s$.}
\label{fig3}
\end{center}
\end{figure}

Finally, we consider non-vanishing masses for both, strange and 
non-strange quarks. Then, according to \eq{masses}, the effective 
Goldstone masses become finite even at leading order, and $m_s$ or $\mu_Q$ 
have to exceed certain threshold values to enable Goldstone boson 
condensation. 
Our results are summarized in \fig{fig3}. The full squares, circles and
triangles correspond to NJL-model calculations with $m_q = 5$~MeV,
10~MeV, and 15~MeV, respectively.   

In the two left figures we show the behavior of the pionic solutions for 
fixed $m_s = 120$~MeV as functions of $\mu_Q$. In the upper panel 
the chiral angle $\theta$ is shown which minimizes the free energy. 
As expected, this angle is zero for small values of $|\mu_Q|$ and becomes
non-zero above a threshold value which depends on the quark mass. 
This behavior can be described well by 
\eqs{masses} to (\ref{delom}) if we treat the constant $a$ which enters
into the expression for the meson masses as a free parameter. 
The dotted lines, which run almost perfectly through the points, have been
obtained with $a = 0.169$. However, this is only about $1/3$ 
of the leading-order result $a = 3\Delta^2/(\pi^2 f_\pi^2)$~\cite{SoSt00},
if we take $\Delta \sim 100$~MeV and $f_\pi = 80$~MeV. 
Again, this discrepancy must be an interaction effect. Nevertheless,
since in \fig{fig2} we found the leading-order results to be in rather
good agreement with the pion channel, a factor 3 in the present
case is a bit surprising. 
To get some insight, we have again repeated the calculation with
$g\Lambda^2 = 1.5$, i.e., $\Delta \sim 30$~MeV.
For this case we find the fitted value for $a$ to be
about $2/3$ of the leading-order prediction. This seems to indicate
that the latter is reached asymptotically, but the convergence is 
relatively slow. A more systematic investigation of this point is 
certainly necessary. 

In the lower left panel we display the free energy gain (for $
g\Lambda^2 = 2.6$). 
The dotted lines indicate the results of \eqs{masses} to (\ref{delom}), 
taking the previously fitted constants $a = 0.169$ and $f_\pi = 80$~MeV.
Obviously, we obtain a perfect description of the NJL model results without 
refitting these constants. 

In the right two figures the analogous quantities are shown for kaon
($K'$) condensates at $\mu_Q = 0$ as functions of the strange quark mass. 
The dotted lines correspond to \eqs{masses} to (\ref{delom}) with the
values for $a$ and $f_\pi$ fitted in the pion channel. 
Obviously, the overall behavior is well reproduced, although the 
quantitative agreement is not as good as in the left two panels. 
In fact, the deviations are consistent with our earlier results:
As one can see in the upper figure, the NJL-model results for 
the thresholds for kaon condensation are at somewhat higher values of $m_s$ 
than the dotted lines suggest. This could again be explained by a 
positive correction term to the kaon mass beyond the leading order. 
Accordingly, the NJL-model results for $\delta\Omega$ (lower right panel)
are also shifted to higher values of $m_s$ as compared with the dotted
lines. On the other hand, for large $m_s$, the light quark masses become
less relevant and the results approach the $m_q = 0$ values of \fig{fig2}.

\section{Summary and outlook}

We have studied pion and kaon condensation in the CFL phase within an
NJL-type model. To that end we have performed a mean-field calculation
allowing for non-vanishing expectation values of certain scalar and
pseudoscalar diquark condensates. 
Main focus of the present article was on the general principle.
We have explicitly shown that Goldstone condensates, i.e., non-vanishing
expectation values of pseudoscalar diquark condensates, develop
in reaction to a finite strange quark mass or a finite
electric charge chemical potential.
In this context it was essential to introduce color chemical potentials
to ensure color neutrality of the CFL and the CFL $+$ Goldstone solutions.

More quantitatively, we found good over-all agreement of the NJL-model
results with predictions obtained within chiral perturbation theory 
in the CFL phase to leading order in the interaction. 
The most intriguing exception is the coefficient $a$, which determines
the dependence of the Goldstone boson mass on the quark masses.
Fitting this coefficient to our results, it turned out to be only
one third of the leading order High Density Effective Theory 
value~\cite{SoSt00} if the coupling is relatively strong 
($\Delta \simeq 100$~MeV) and two thirds for a weaker coupling 
($\Delta \simeq 30$~MeV). The leading-order value may thus be reached
at very weak couplings only, but this deserves further investigations.
Apart from this point we found some deviations in the kaon sector which
we also identified as interaction corrections to the leading order.

Having demonstrated the general consistency of the method, we should now
extend our analysis to perform a more complete study of the phase
structure of strongly interacting matter at high densities. 
In the present article, we have in most cases restricted the space of 
possible diquark condensates to certain chiral rotations, in order to 
facilitate the comparison with the effective theories. This restriction
should be relaxed. As shown in \fig{fig1}, this can lead to a further 
reduction of the free energy.  

The model should also be extended to include quark-antiquark
condensates, like $\ave{\bar uu}$, $\ave{\bar dd}$, and $\ave{\bar ss}$,
which lead to density dependent effective quark masses 
and in this way could influence the phase structure 
considerably~\cite{BuOe02}.
Now, in order to study Goldstone boson condensation, we also need to take into
account the corresponding chiral partners, $\ave{\bar q\,i\gamma_5 \tau_a q}$.
These condensates are the essential degrees of freedom to describe pion 
and kaon condensates in color non-superconducting phases, which have 
recently been studied in great detail in Ref.~\cite{BCPR04}. 
In the CFL phase, the dominant effects should be due to the diquark 
condensates, but nevertheless, quark-antiquark condensates could lead to
important corrections. 
We should also note that our toy-model Lagrangian, \eq{lagrangian},
misses instanton effects, which could significantly contribute 
to the Goldstone boson masses~\cite{MaTy00}. 
Finally, it would be interesting to extend the analysis to the
gapless CFL phase.
\\[3mm]

{\bf Acknowledgments:}\\
I am particularly grateful to Thomas Sch\"afer for many clarifying 
comments.
I also thank Micaela Oertel, Krishna Rajagopal, Sanjay Reddy,  
Bernd-Jochen Schaefer, and Igor Shovkovy
for valuable comments and discussions.
Related work has been done independently by M. Forbes~\cite{mforbes}. 
I also thank him for the subsequent communication 
which stimulated the study of the $\Delta$ dependence of the
results in the revised manuscript.


\begin{thebibliography}{99}
\bibitem{ARW99}  M. Alford, K. Rajagopal, and F. Wilczek,
                 Nucl. Phys. B 537 (1999) 443.
\bibitem{Sho99}  I.A. Shovkovy and L.C.R. Wijewardhana,
                 Phys. Lett. B 470 (1999) 189.
\bibitem{Sch00b} T. Sch\"afer, Nucl. Phys. B 575 (2000) 269.
\bibitem{EHHS00} N. Evans, J. Hormuzdiar, S.D. Hsu, and M. Schwetz,
                 Nucl. Phys. B 581 (2000) 391.
\bibitem{RaWi00} K. Rajagopal and F. Wilczek,
                 ``The Condensed Matter Physics of QCD'', 
                 in: B.L. Ioffe Festschrift \emph{At the Frontier of
                 Particle Physics / Handbook of QCD}, vol. 3,
                 edited by M. Shifman, 
                 World Scientific, Singapore, 2001, pp.~2061--2151.
\bibitem{Alford} M. Alford, Ann. Rev. Nucl. Part. Sci. 51 (2001) 131.
\bibitem{Schaefer} T. Sch\"afer, hep-ph/0304281,
                 to appear in the proceedings of the BARC workshop.
\bibitem{Rischkereview} D.H. Rischke, Prog. Part. Nucl. Phys. 52 (2004) 197. 
\bibitem{Sch00c} T. Sch\"afer, Phys. Rev. Lett. 85 (2000) 5531.
\bibitem{BS02}   P.F. Bedaque and T. Sch\"afer, Nucl. Phys. A 697 (2002) 802.
\bibitem{KR02}   D.B. Kaplan and S.Reddy, Phys. Rev. D 65 (2002) 054042.
\bibitem{SoSt00} D.T. Son and M.A. Stephanov, Phys. Rev. D 61 (2000) 074012;
                 erratum ibid D 62 (2000) 059902. 
\bibitem{CaGa99} R. Casalbuoni and R. Gatto, Phys. Lett. B 464 (1999) 111.
\bibitem{CGN01}  R. Casalbuoni, R. Gatto, and G. Nardulli,
                 Phys. Lett. B 498 (2001) 179;
                 erratum ibid B 517 (2001) 483. 
\bibitem{Hong00} D.K. Hong, Phys. Lett. B 473 (2000) 118.
\bibitem{BBS00}  S.R. Beane, P.F. Bedaque, and M.J. Savage,  
                 Phys. Lett. B 483 (2000) 131.
\bibitem{buballa} M. Buballa, hep-ph/0402234, Phys. Rep. in print.
\bibitem{BCPR04} A. Barducci, R. Casalbuoni, G. Pettini, and L. Ravagli,
                 hep-ph/0410250.
\bibitem{AKR04}  M. Alford, C. Kouvaris, and K. Rajagopal,
                 Phys. Rev. Lett. 92 (2004) 222001; hep-ph/0406137
                 to appear in Phys. Rev. D.
\bibitem{RSR04}  S.B. R\"uster, I.A. Shovkovy, and D.H. Rischke,
                 Nucl. Phys. A 743 (2004) 127.
\bibitem{FKR04}  K. Fukushima, C. Kouvaris, and K. Rajagopal,
                 hep-ph/0408322.
\bibitem{BuOe02} M. Buballa and M. Oertel, Nucl. Phys. A 703 (2002) 770.
\bibitem{SRP02}  A. Steiner, S. Reddy, and M. Prakash, 
                 Phys. Rev. D 66 (2002) 094007.
\bibitem{Kry03}  A. Kryjevski, Phys. Rev. D 68 (2003) 074008.
\bibitem{KKS04}  A. Kryjevski, D. Kaplan, and T. Sch\"afer, hep-ph/0404290.
\bibitem{MaTy00} C. Manuel and M.H. Tytgat, Phys. Lett. B 479 (2000) 190.
\bibitem{mforbes} M.M. Forbes, hep-ph/0411001.
\end{thebibliography}
\end{document}